\documentclass[10pt,twocolumn,letterpaper]{article}

\usepackage{cvpr}      

\usepackage[dvipsnames]{xcolor}

\usepackage{float}

\definecolor{cvprblue}{rgb}{0.21,0.49,0.74}
\usepackage[pagebackref,breaklinks,colorlinks,citecolor=cvprblue]{hyperref}

\newcommand\blfootnote[1]{%
  \begingroup
  \renewcommand\thefootnote{}\footnote{#1}%
  \addtocounter{footnote}{-1}%
  \endgroup
}

\def\paperID{*****} 
\def\confName{CVPR}
\def\confYear{2024}

\title{Alljoined1 - A dataset for EEG-to-Image decoding}

\author{Jonathan Xu \textsuperscript{* 1 2}, Bruno Aristimunha \textsuperscript{* 3 4}, Max Emanuel Feucht \textsuperscript{* 1 5}, \\ Emma Qian \textsuperscript{\textdagger 1}, Charles Liu \textsuperscript{\textdagger 1 2},  Tazik Shahjahan \textsuperscript{\textdagger 1 2}, Martyna Spyra \textsuperscript{\textdagger 1},  Steven Zifan Zhang$^{1, 6}$,\\ Nicholas Short$^{1, 6}$, Jioh Kim$^{1, 6}$, Paula Perdomo$^{1, 6}$, Ricky Renfeng Mao$^{1, 2}$, Yashvir Sabharwal$^{1}$, \\Michael Ahedor$^{1}$, Moaz Shoura$^{6}$, Adrian Nestor$^{6}$
}

\begin{document}
\maketitle

\blfootnote{\noindent \textsuperscript{*} equal contribution. \textsuperscript{\textdagger} core contribution. $^1$Alljoined $^2$The University of Waterloo $^3$Université Paris-Saclay, Inria TAU, CNRS, LISN $^4$Federal University of ABC $^5$Vrije Universiteit Amsterdam $^6$University of Toronto}

\begin{abstract}
We present Alljoined1, a dataset built specifically for EEG-to-Image decoding. Recognizing that an extensive and unbiased sampling of neural responses to visual stimuli is crucial for image reconstruction efforts, we collected data from 8 participants looking at 10,000 natural images each. We have currently gathered 46,080 epochs of brain responses recorded with a 64-channel EEG headset. The dataset combines response-based stimulus timing, repetition between blocks and sessions, and diverse image classes with the goal of improving signal quality. For transparency, we also provide data quality scores. We publicly release the dataset and all code at \href{https://linktr.ee/alljoined1}{https://linktr.ee/alljoined1}.
\end{abstract}    
\section{Introduction}
\label{sec:intro}

In the fields of cognitive neuroscience and medical imaging, advancements in deep learning have led to unparalleled precision in decoding brain activity \cite{Roy2019, jayaram2018moabb, scotti2024mindeye, scotti2023reconstructing, benchetrit2024brain}. 
Researchers have translated the intricate patterns of brain activity during various cognitive processes by utilizing neuroimaging modalities, such as functional Magnetic Resonance Imaging (fMRI) and electroencephalography (EEG).

In this context, one particular area of interest is image reconstruction, which involves the decoding of neural responses to visual stimuli, offering insights into how the brain encodes and processes visual information \cite{radford2021learning, scotti2023reconstructing, benchetrit2024brain, king:2020, chenstructure, chen2024cinematic}.

While fMRI has traditionally been the primary tool for image reconstruction due to its excellent spatial resolution, its low temporal resolution severely delimits actual clinical usage. 
On the other hand, EEG is a medical modality available in everyday clinical contexts with an excellent time resolution \cite{nemrodov2018neural, singh2024learning, singh_eeg2image_2023}. 
As neurons fire at millisecond scales, the high temporal resolution provided by EEG is crucial for real-time monitoring of neural dynamics \cite{thorpe1996speed, dijkstra2018differential, harel2016temporal}. 
Additionally, EEG is portable, more accessible to set up, and much more cost-effective than fMRI, making it suitable for real-world applications, including brain-computer interfaces and clinical diagnostics.

The development of very large fMRI-to-image datasets has proven foundational for recent breakthroughs in deep-learning image reconstruction projects. 
Inspired by the need for such datasets in the EEG domain, we present \textbf{Alljoined1}, a novel, large-scale dataset covering a wide range of naturalistic stimuli that allows for robust, generalizable image reconstruction efforts. Our contributions are as follows:

\begin{itemize}
    \item We propose a stimulus presentation approach that tailors trial duration and session and block repetitions to maximize the signal-to-noise (SNR) ratio.
    \item We introduce a diverse dataset of EEG responses to 9k unique naturalistic images for each of the eight participants, with 1k additional images shared between participants.
    \item We perform qualitative comparisons against current EEG-to-image datasets.
\end{itemize}

\section{Related Work}
\label{sec:related}

\subsection{EEG-to-Image Datasets} 

EEG-to-image datasets consist of EEG waveforms recorded while participants watch visual stimuli, enabling the study of neural representations in the brain. However, previous research on EEG-based image reconstruction has often relied on datasets exhibiting severe limitations regarding acquisition design or generalizability to naturalistic stimuli \cite{wakita_photorealistic_2021, le2021brain2pix, singh_eeg2image_2023}.

A popular EEG-image dataset is \emph{Brain2Image} \cite{kavasidis_brain2image_2017}, which consists of evoked responses to a visual stimulus from distinct image classes. Each block consists of stimuli corresponding only to a single image class. There are 40 classes, with 50 unique images in each class. This dataset has been criticized for having no train-test separation during recording, block-specific stimuli patterns, and lack of consistency across different frequency bands. These factors can incorrectly boost model performance by giving extraneous proxy information about the block rather than the actual image-specific brain responses \cite{9264220, ahmed_confounds_2022}. An extensive study highlights how many recent EEG-based image reconstruction attempts depend crucially on their block design, demonstrating how similar analytical approaches are not capable of meaningfully decoding EEG signals in a rapid serial visual paradigm (RSVP) \cite{9264220}, even when collecting large amounts of data for only a single subject \cite{9578178}.

Recent studies achieving impressive reconstruction results have relied on data collected with flawed block designs \cite{lan_seeing_2023, bai2023dreamdiffusion, khaleghi_visual_2022}, calling the validity of their results into question. As recommended by \cite{9264220, ahmed_confounds_2022}, the stimuli within each block in our dataset were chosen randomly across a variety of natural images, effectively minimizing the risk of block-class correlations.

The diversity of decoding stimuli further limits current EEG-based image reconstruction datasets. While studies like \textit{Brain2Image} or \cite{9578178} consist of images belonging to 40 classes, several studies utilize a dataset of visual imagery of characters and objects belonging to only 10 different classes, \textit{ThoughtViz} \cite{tirupattur_thoughtviz_2018}. Such a discretized representation of real world objects fails to account for the continuous, diverse quality of naturalistic stimuli. The same limitation applies to studies utilizing a severely limited quantity of naturalistic stimuli. Approaches to EEG-based image reconstruction derived from the \textit{ThoughtViz} \cite{mishra_neurogan_2023, singh_eeg2image_2023}, \textit{Brain2Image} \cite{lan_seeing_2023, bai2023dreamdiffusion, khaleghi_visual_2022}, or other equally selective datasets \cite{ahmadieh2024visual, wakita_photorealistic_2021}, may thus suffer from generalizing well to diverse, real-world stimuli. Moreover, a limited number of image classes may encourage image reconstruction models to generate images class-conditionally, rather than reconstructing images based on (continuous) brain-encoded semantic or perceptual attributes of an image.

To account for the diverse and continuous nature of naturalistic images, Alljoined1 consists of 1) 10,000 images per participant 2) that belong to at least one of 80 MS-COCO \cite{lin2014microsoft} object categories. Importantly, each MS-COCO category is broader than a single object class (e.g. the \emph{things} category includes car, skateboard, hat, etc.), and each image can belong to up to 5 classes \cite{coco2014}.

There are also existing datasets that include naturalistic stimuli, but compromise in other domains. The \textit{MindBigData} initiative \cite{vivancos2022mindbigdata}, or \cite{9578178} capture a wide sample of images, but are derived only from a single individual, limiting the potential of training image reconstruction models that generalize to other individuals. The \textit{THINGS-EEG1} \cite{grootswagers2022human} and \textit{THINGS-EEG2} \cite{gifford_large_2022} datasets were acquired using short image presentation times of 50 and 100 ms, and a stimulus onset asynchrony of 100 and 200 ms. 

Although the rapid serial visual presentation \cite{grootswagers2019representational} paradigm proposes disentangling the temporal dynamics of visual processing and categorical abstraction of non-target stimuli, it is not ideal for capturing cortical image processing beyond early visual activity with low noise. We see that \citep{spampinato2017deeplearning} obtained the highest accuracy with their EEG-image classifier when focusing on 320-480 ms after stimulus onset, and \cite{nemrodov2020multivariate} is able to extract relevant decoding features even around 550 ms after stimulus onset. This suggests that while it takes 50-120 ms for object recognition of a stimulus to register in the visual cortex, a longer stimulus period is beneficial for accuracy on downstream tasks. Alljoined1 consists of extensive data from eight participants, measured with an inter-stimulus interval of 300 ms, which captures important hallmarks in visual processing while maintaining a high presentation frequency \cite{ teichmann2023multidimensional, grootswagers2022human}. This setup might furthermore allow us to overcome the limitations in decoding image content from EEG activity in RSVP paradigms, as previously reported in \cite{9578178}. 

\subsection{fMRI-Image Datasets}

The recent development of large functional magnetic resonance imaging (fMRI) datasets has enabled researchers to decode and reconstruct images observed by humans with unprecedented accuracy. 

The \textit{Brain, Object, Landscape Dataset} (BOLD5000) \cite{chang2019bold5000} contains brain responses from 4 human participants who viewed 5,254 images depicting natural scenes from the Scene UNderstanding (SUN) \cite{yamins2016using}, MS-COCO \cite{lin2014microsoft}, and ImageNet datasets  \cite{deng2009imagenet}. Similarly, in the \textit{Generic Object Decoding Dataset} (GOD) \cite{horikawa2017god}, 1,200 images from the ImageNet database were cropped and shown to 5 participants, resulting in one of the first datasets to establish methods for decoding generic object categories from brain activity. 

The \textit{Natural Scenes Dataset} (NSD) \cite{allen2022massive} consists of the brain responses of 8 human participants passively viewing 9,000–10,000 color natural scenes from MS-COCO. This magnitudes-larger dataset has fueled leaps in reconstruction accuracy seen in recent work like MindEye2 \cite{scotti2024mindeye}. However, the adaptation of such impressive achievements to real-life contexts is quite limited, as MRI scanners are notoriously expensive and difficult to access.
\newcommand{\X}{\mathcal{X}\xspace}
\newcommand{\Y}{\mathcal{Y}}
\newcommand{\x}{\mathbf{x}}
\newcommand{\y}{\mathbf{y}}

\section{Methods and Materials}
\label{sec:method}

\subsection{Participants}

We collected data from eight participants (six male, two female), with an average age of 22 $\pm$ 0.64 years, all with normal or corrected-to-normal vision, right-handed. All participants were healthy, with no neurocognitive impairments, except 2 participants who reported a history of mental health disorders (e.g. GAD, ADHD). Each participant provided informed consent. The Research Ethics Board approved the procedures as \textbf{suppressed for double-blind review}. We note that there are potential limitations of the study due to the imbalance between the genders of the participants and the low age disparity, which could influence bias and learning with AI models.
\subsection{Stimuli} \label{sec:sti}
We use the same visual stimuli as what was shown in the fMRI Natural Scenes Dataset (NSD) \cite{allen2022massive}, consisting of $70,566$ images portraying everyday objects and situations in their natural context. All NSD images are drawn from the MS-COCO dataset \cite{coco2014}, including annotations about objects and their corresponding category contained in the image. Each image can contain more than one object and more than one object category. These fine-grained object categories are further grouped into \emph{supercategories}, each of which comprehensively includes all related categories as defined subsets. 

\begin{figure}[H]
    \centering
    \includegraphics[width=0.45\textwidth]{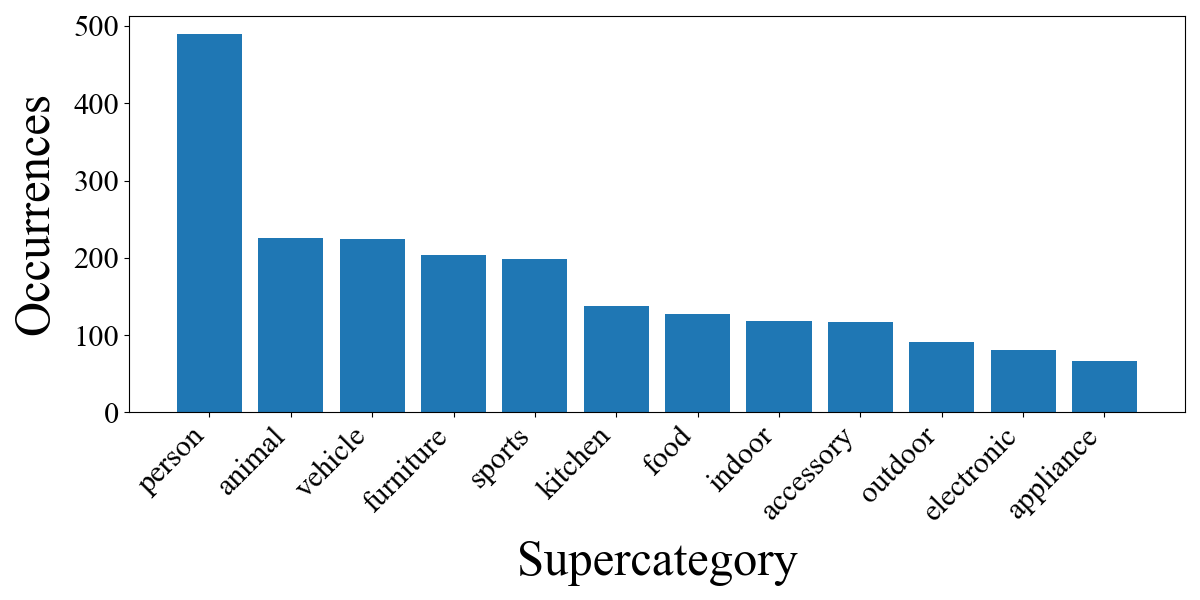}
    \caption{Top 12 most frequently occurring \emph{supercategories} in our dataset.}
    \label{fig:supercategory}
\end{figure}

The current study uses a subset of the first $960$ images in the $1000$ images shown across all participants in the NSD study. 
These images are drawn from the \emph{shared1000} subset of the NSD dataset, which comprises 1000 specially curated images that all participants in the original NSD study were presented with \cite{allen2022massive}. Within this subset of the NSD dataset, the supercategory \emph{person} was most represented, occurring in 50.94\% of all images, followed by animal (23.54\%) and vehicle (23.33\%). 
The distribution of the supercategories is shown in \autoref{fig:supercategory}.

\subsection{Procedure}

Images were displayed to participants over the course of multiple one-hour-long sessions. Each session consisted of 16 blocks, wherein images in the first 8 unique blocks were repeated in the second 8 blocks. 

The repeated blocks (e.g., blocks 1 \& 8, 2 \& 9, etc.) contained the same stimuli but in a shuffled order to avoid sequence effects. 
Within each block, 120 images from the NSD dataset were presented twice, as well as 24 oddball stimuli, amounting to 264 images per block. 

Given the within-block and the between-block repetitions of NSD images, each NSD image was presented 4 times to obtain a higher signal-to-noise ratio of the evoked neural responses. 
Within each trial, an image (NSD or oddball) was presented for $300$ ms, followed by $300$ ms of a black screen; a white fixation cross was visible on the screen throughout the entire trial. 

At the end of each trial, an extra jitter time between 0-50 ms was added for randomness. 
To ensure focus, participants were prompted to press the space bar when two consecutive trials contained the same image. These oddball trials occurred 24 times within each block; oddballs trials have been discarded from the dataset due to motion artifacts, EEG repetition suppression, and other issues.

\begin{figure*}
    \centering
    \includegraphics[width=\linewidth]{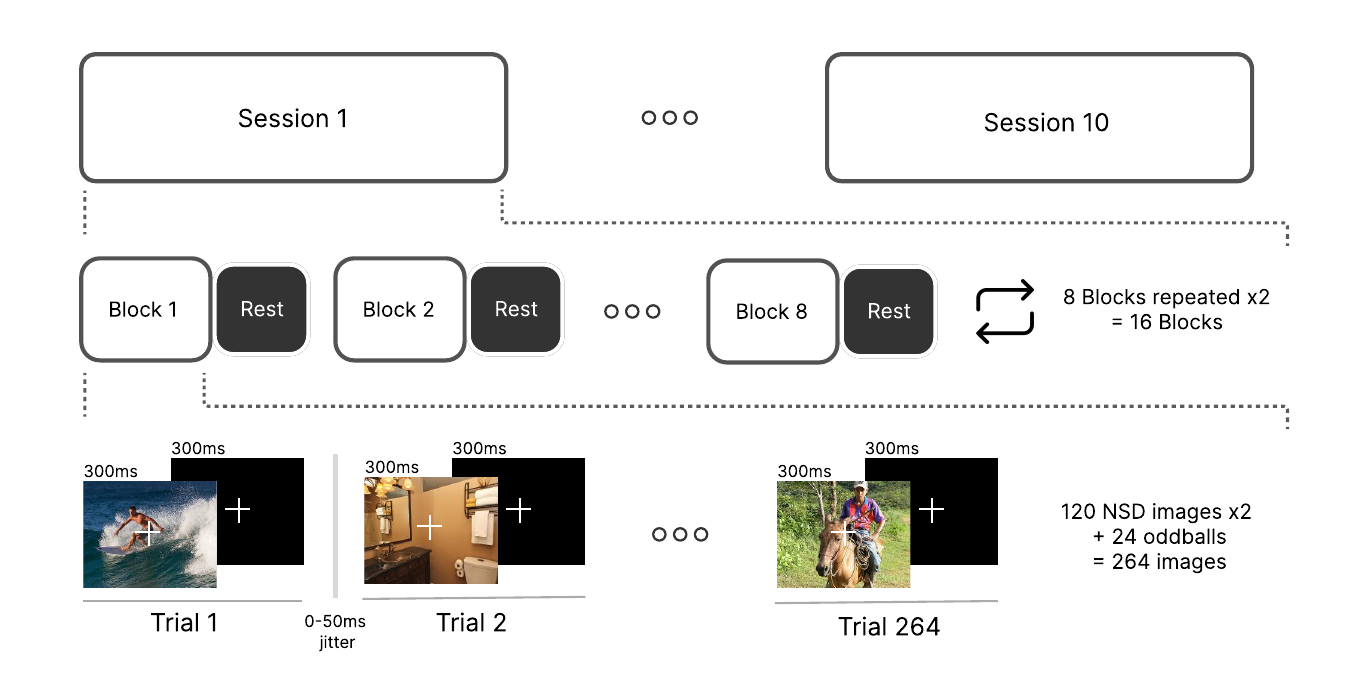}
    \caption{Schematic overview of the structure of trials, blocks, and sessions. Each of the 120 block-specific NSD images is presented twice within each block, and each of the 8 session-specific blocks is presented twice within each session. Each participant performed two sessions on different days. Each of the 10 sessions thus consists of 960 NSD images repeated four times within and across blocks, totaling 9600 unique NSD images per participant.}
    \label{fig:diagram-block-sessions}
\end{figure*}


\subsection{Hardware Setup}

\begin{figure}[!ht]
    \centering
    \includegraphics[width=0.45\textwidth]{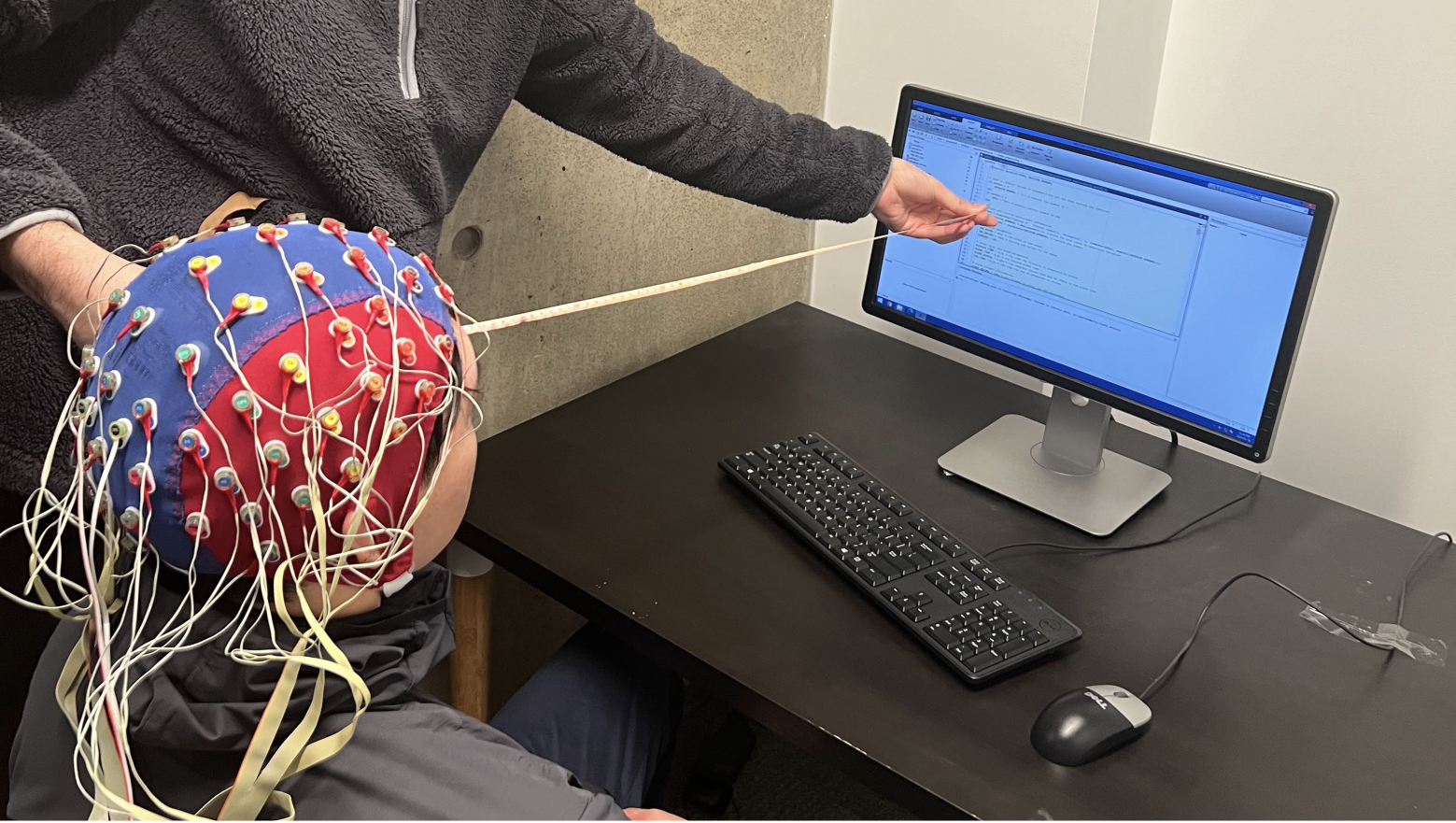}
    \caption{Experimental setup with monitor 80 cm from participant.}
    \label{fig:monitor}
\end{figure}

We recorded data using a 64-electrode BioSemi ActiveTwo system, digitized at a rate of $512$ Hz with 24-bit A/D conversion. The montage was arranged in the International 10-20 System, and the electrode offset was kept below 40 mV. We used a $22$ inch Dell monitor at a resolution of 1080p/60Hz to display the visual stimulus. As depicted in \autoref{fig:monitor}, the monitor was positioned centrally and placed at a distance of $80$ cm to maintain a $3.5$° visual angle of stimuli. We avoided larger angles to minimize the occurrence of gaze drift.

\subsection{Pre-processing}

Regarding the dataset pre-processing, we follow recent work on the importance of separating the biomarkers from the central nervous and peripheral systems, as described in \cite{Bomatter2023}, and applied the minimum necessary steps. This dataset was pre-processed using the \textsc{MNE-Python} library \cite{gramfort2013meg}.

\paragraph{Filtering}

Initially, we applied band-pass filtering with a low frequency of $0.5$ Hz and a high frequency of $125$ Hz with overlap-add finite impulse response filtering, with range based on \cite{tao2021gated}. We then apply a notch filter at 60Hz to eliminate power line noise.

\paragraph{Independent Component Analysis (ICA)} 
Next, we performed an ICA decomposition using a FastICA model \cite{hyvarinen2001independent, ablin2018faster} to separate non-gaussian biological artifacts noise from the signal source. We used a decomposition that retained 95\% of the variance and excluded ICs corresponding to eye blinks on the raw data.

\paragraph{Epoching} 
We segmented the continuous data into \textit{epochs}, with each epoch starting at $-50$ ms onset stimulus and ending at the end of each trial at $600$ ms, as described in Section \ref{sec:sti}. The inter-trial jitter periods were excluded from the epochs.

 
\paragraph{Artifact correction}

We used  the \textsc{Autoreject} algorithm \cite{autoreject2017jas} to identify and handle artifact-heavy epochs. Autoreject employs a peak-to-peak threshold criterion separately for each sensor to determine whether an epoch should be (i) repaired by interpolating the affected sensors using neighboring sensors, or (ii) entirely excluded from further analysis. It performs grid search to determine appropriate values for $\rho$, the number of channels to interpolate, and $\kappa$, the percentage of channels that must agree as a fraction of total channels for consensus. By looking at the number of erroneous sensors per trial, this approach allows correction on a per-trial basis instead of applying a single global threshold to all trials. A mean of 130.75 epochs was dropped per session, with a standard deviation of 260.44.

\paragraph{Baseline correction}

Finally, we re-reference our channels using an average reference scheme, before applying a baseline correction window from $-50$ ms to $0$ ms relative to stimulus onset, following recommendations from \cite{tanner2016high} for ERP baseline. The epoch data subtracts the average activation during the baseline interval to remove noise from the signal.

\section{Analysis}
\label{sec:analysis}

\begin{figure*}
\centering
    \begin{subfigure}{0.49\linewidth}
    \includegraphics[width=\linewidth]{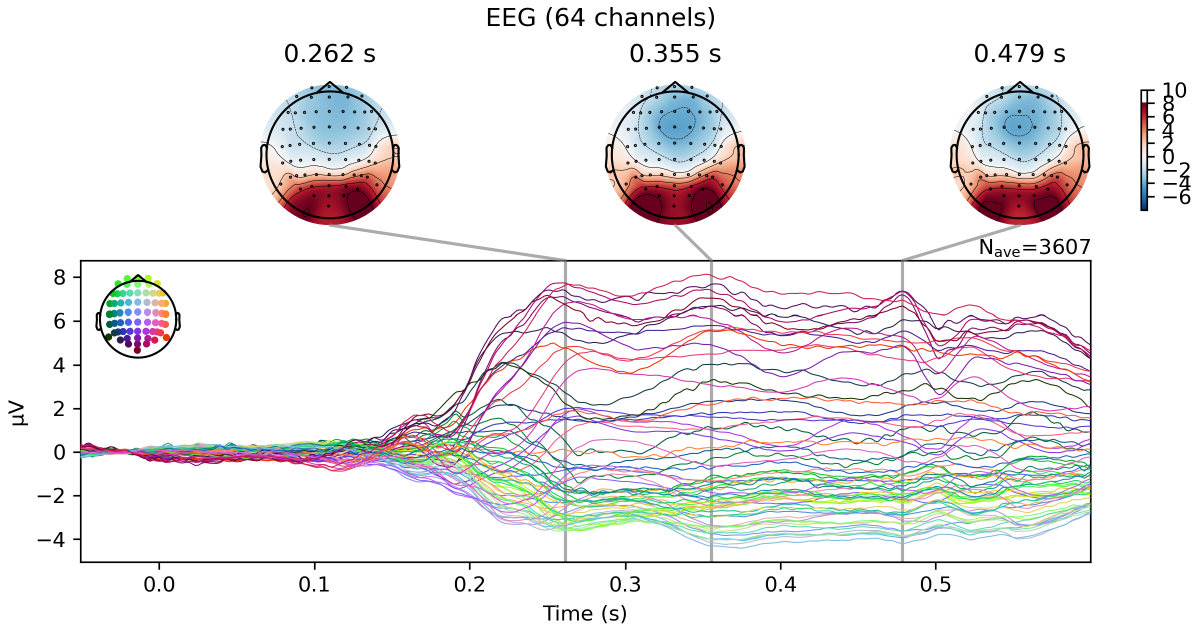}
    \caption{First session of the fifth participant}
    \label{fig:short-a}
  \end{subfigure}
  \begin{subfigure}{0.49\linewidth}
    \includegraphics[width=\linewidth]{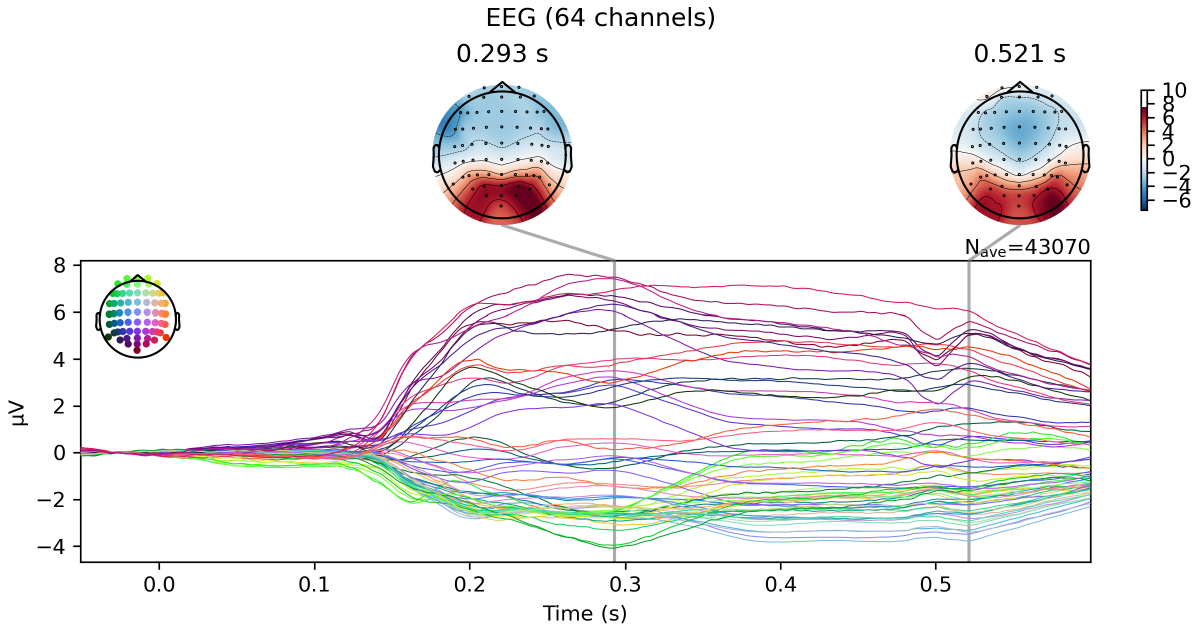}
    \caption{All sessions across all participants}
    \label{fig:short-b}
    \end{subfigure}
  \caption[width=.8\linewidth]{EEG topographic maps and corresponding signals at all 64 electrodes averaged over a) 3823 events for the fifth participant (left) and b) across all sessions for all participants (43070 events) in the Alljoined1 dataset (right), highlighting individual and common brain activity patterns associated with image presentation. An \textit{event} is defined as a specific time point in the experiment.}
  \label{fig:subject_level_eeg}
\end{figure*}

\subsection{ERP Analysis}
\begin{figure}
    \centering
    \includegraphics[width=0.45\textwidth]{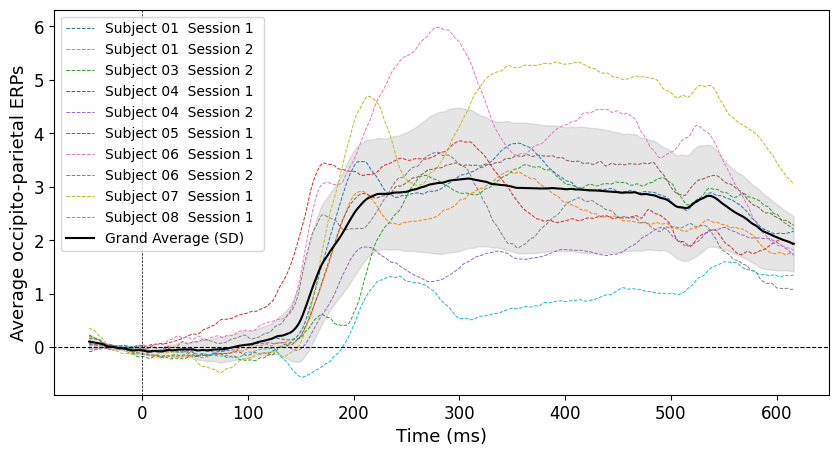}
    \caption{ERPs averaged over occipital and parietal electrodes for all participants and sessions. Shaded areas around the grand average ERP indicate standard deviations at all timepoints.}
    \label{fig:averaged-erp}
\end{figure}

The distribution of event-related potentials (ERPs) across all 64 channels is displayed for a single session of one participant and averaged over all participants and sessions in \autoref{fig:averaged-erp}. We observe a strong consistent rise in activity beginning after 150 ms, with a peak between 250 and 300 ms. Note that this latency aligns with the timing parameters of our experimental design, which involves a 300ms presentation followed by a 300ms rest period, with an additional 0-50ms jitter. Both participant- and cohort-level activity exhibits a sustained high level of activity up until 500 ms after stimulus onset, where a consistent dip in activity is observed for both the single participant as well as the whole cohort. 

Topographies of activation additionally reveal a strong concentration of positive activation at occipital, parietal, and partially temporal electrode locations and a consistently negative activation at central and frontal areas. This topographical distribution was stable across the duration of the ERP and corresponded well between the single participant and the cohort. Given the strong peak in activity at the occipital and parietal areas, we further investigated the distribution of ERPs across individual participants and sessions at the occipital and parietal electrodes, as displayed in figure \ref{fig:averaged-erp}. While the magnitude of activation differs between participants, we conclude a by-and-large consistent activation pattern across participants and sessions. 

\subsection{SNR Analysis}

\begin{figure*}
    \centering
  \includegraphics[width=1\textwidth]{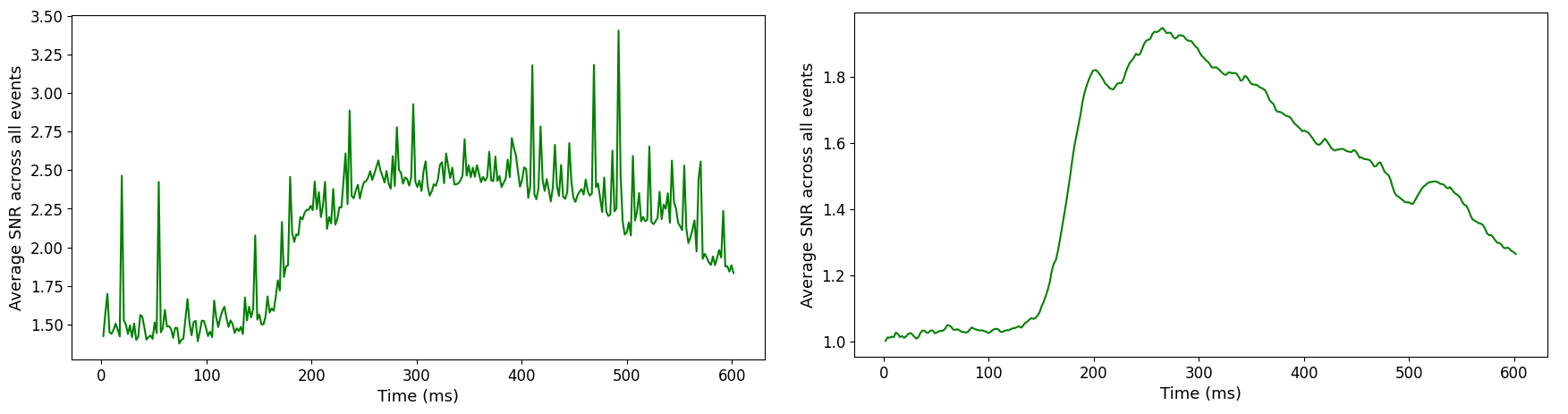}
  \caption[width=.8\linewidth]{Signal to Noise Rate (SNR) averaged across each session, across each block, and within each block for participant 5. Left: SNR for only the first session 1, Right: SNR for all sessions.}
  \label{fig:snr}
\end{figure*}

The Signal-to-Noise Ratio (SNR) serves as a pivotal metric in evaluating the efficacy of our dataset. To ascertain the SNR, we employ the Standardized Measurement Error (SME) as a gauge for noise assessment \cite{luck2021standardized}. We choose SME as our metric of choice as it is able to robustly quantify the data quality for each participant at each electrode site \cite{luck2020blog}. The SME is determined by calculating the standard deviation of the aggregated waveform average for each event type across all trials and then dividing this by the square root of the event type's occurrence count. The SNR is subsequently derived by dividing the mean signal values by their corresponding SME.

\autoref{fig:snr} compares the average SNR across all events in a single session for participant 5 with the average SNR across all events for both sessions concatenated. We see that the SNR is noticeably lower in the multi session graph. This is due to the increased number of repetitions for a given event at different timepoints. This leads to a disproportionately higher standard deviation value and consequently a higher SME and lower SNR. However, that is not to say that the quality of the data is worse. It actually reflects more accurate SNR values as there are more data points, distributed across different sessions. It is also observed that the single sessions graph is more volatile across time, demonstrating a greater variance in SNR values which are captured by having a less accurate metric for noise with less trials to average between. This fluctuation underscores the limited accuracy of noise metrics derived from fewer trials, thus highlighting the critical importance of incorporating repeated measures across sessions or blocks for robust SNR evaluation.

Furthermore, we observe a strong SNR increase 150 ms after stimulus onset. Note that this increase in SNR exhibits the same spiking timing we see in our earlier topographic maps and averaged ERP graphs, suggesting that meaningful activity starts to surface with a considerable delay with respect to stimulus onset.

\subsection{Discussion}

The ERP and topography analyses, as well as our analysis of the SNR reveal and reinforce several benefits of the acquired dataset, with regard to the stimulus design and timing. 

\begin{enumerate}
    \item \textbf{Stimulus duration:} A 300 ms presentation window as well as a subsequent 300 ms rest period allows the capture of both early and late cognitive processes, as evidenced by the single subject peaks at around 262 ms up to 479 ms in \autoref{fig:snr} a), and the averaged peaks at 293 ms and 521 ms in \autoref{fig:snr} b), respectively. The duration of 300 ms for image presentation is sufficient for the brain to engage in both perceptual encoding and initial stages of memory processing, which may not be as effectively captured with shorter presentation times. The subsequent 300ms rest period provides a window to measure the brain's higher-level visual and semantic response to the stimuli. The whole ERP thus not only reflects the initial feed-forward transfer of sensory information to visual cortical areas but also the subsequent recurrent interactions involved in attention and semantic analysis, that unfold over hundreds of milliseconds after stimulus onset.
    The relevance of longer presentation times and longer stimulus-onset asynchrony is additionally supported by the sustained ERP activation presented in \autoref{fig:averaged-erp}, as well as the latency of SNR increase and peak in \autoref{fig:snr}.

    \item \textbf{Comparison with prior studies:} Presentation times of only 100 ms, or stimulus onset asynchronies of only 200 ms fail to capture the rich neural dynamics associated with image processing, involving both lower and higher level processing. In THINGS EEG2 \cite{gifford_large_2022}, with a shorter 100ms presentation time followed by 100ms of rest, the stimulus exposure may have been insufficient to elicit the full range of cognitive processes to occur. The limited time window could explain the lesser degree of neural activity in the corresponding time window. Similarly, THINGS EEG1 \cite{grootswagers2022human} employed a shorter 50ms presentation window followed by 50ms rest, which, while suitable for examining the earliest stages of sensory processing in the visual cortex, likely precluded the phases of the cognitive processes that unfold over a longer period. This includes higher-order mechanisms such as selective attention, working memory updating, and retrieval of semantic associations from long-term memory stores \cite{ku2018selective}.

    \item \textbf{Phase locking mitigation:}
    The inclusion of a jitter ranging from 0-50ms helps mitigate phase locking, a phenomenon where the participant's alpha-wave activity becomes synchronously aligned with the pattern of the stimuli after repeated presentations.
    
    \item \textbf{Anticipatory bias minimization:} Additionally, the jitter prevents the participants from predicting the exact onset of the next stimulus, thus reducing the potential for anticipatory neural activity that could confound the data.
    
\end{enumerate}

In conclusion, we choose a 300ms latency as it provides a good trade-off between capturing long-term neural activity whilst maintaining a high presentation frequency. The ideal timing of our experiment ensures the acquisition of a comprehensive ERP waveform, contributing to a more nuanced understanding of cognitive processes and neural dynamics as compared to the shorter intervals used in THINGS EEG2 and THINGS EEG1.

\section{Conclusion}
\label{sec:conclusion}

We introduce Alljoined1, an EEG-image dataset that uses well-timed stimuli, repetitions between blocks and sessions, and a wide distribution of natural images to create an improved dataset for image decoding tasks. We believe that its size, diversity, and quality will help promote work to better understand the mechanisms of visual processing, and in decoding visual responses in clinical and consumer brain-computer interface (BCI) contexts.

\textbf{Future directions:} We are eager to explore high-density EEG recordings of exclusively the occipital and parietal regions to better target regions of the brain most responsive to visual stimuli. We are also interested in conducting ablation studies on the generalizability of responses to imagined mental imagery. We further believe there is great potential in exploring continuous data collection in natural environments with a wireless headset.

\textbf{Data availability:} Both the raw and preprocessed EEG dataset is available on \href{https://osf.io/kqgs8/?view_only=3274f4ce973e4240b01df955445daad6}{OSF}. Labels to the corresponding NSD image IDs are included in the object files.

\textbf{Code availability:} The stimulus and preprocessing code to reproduce all the results is available on Anonymous GitHub \href{https://anonymous.4open.science/r/alljoined1-stimulus}{here} and \href{https://anonymous.4open.science/r/alljoined1_preprocessing}{here}.

\section{Acknowledgements}

This work was sponsored by Z Fellows, Hack Grants, Moth Fund and Fiona Leng.Bruno's work was supported by DATAIA Convergence Institute as part of the ``Programme d’Investissement d’Avenir'', (ANR-17-CONV-0003) operated by LISN-CNRS. 
We would like to thank Dr Sylvain Chevallier for his valuable feedback on this manuscript.

{
    \small
    \bibliographystyle{ieeenat_fullname}
    \bibliography{main}

\begin{thebibliography}{51}
\providecommand{\natexlab}[1]{#1}
\providecommand{\url}[1]{\texttt{#1}}
\expandafter\ifx\csname urlstyle\endcsname\relax
  \providecommand{\doi}[1]{doi: #1}\else
  \providecommand{\doi}{doi: \begingroup \urlstyle{rm}\Url}\fi

\bibitem[Ablin et~al.(2018)Ablin, Cardoso, and Gramfort]{ablin2018faster}
Pierre Ablin, Jean-Fran{\c{c}}ois Cardoso, and Alexandre Gramfort.
\newblock {Faster ICA under orthogonal constraint}.
\newblock In \emph{2018 IEEE International Conference on Acoustics, Speech and Signal Processing (ICASSP)}, pages 4464--4468. IEEE, 2018.

\bibitem[Ahmadieh et~al.(2024)Ahmadieh, Gassemi, and Moradi]{ahmadieh2024visual}
Hajar Ahmadieh, Farnaz Gassemi, and Mohammad~Hasan Moradi.
\newblock {Visual image reconstruction based on EEG signals using a generative adversarial and deep fuzzy neural network}.
\newblock \emph{Biomedical Signal Processing and Control}, 87:\penalty0 105497, 2024.

\bibitem[Ahmed et~al.(2021)Ahmed, Wilbur, Bharadwaj, and Siskind]{9578178}
Hamad Ahmed, Ronnie~B. Wilbur, Hari~M. Bharadwaj, and Jeffrey~Mark Siskind.
\newblock Object classification from randomized eeg trials.
\newblock In \emph{2021 IEEE/CVF Conference on Computer Vision and Pattern Recognition (CVPR)}, pages 3844--3853, 2021.

\bibitem[Ahmed et~al.(2022)Ahmed, Wilbur, Bharadwaj, and Siskind]{ahmed_confounds_2022}
Hamad Ahmed, Ronnie~B. Wilbur, Hari~M. Bharadwaj, and Jeffrey~Mark Siskind.
\newblock {Confounds in the {Data}—{Comments} on “{Decoding} {Brain} {Representations} by {Multimodal} {Learning} of {Neural} {Activity} and {Visual} {Features}”}.
\newblock \emph{IEEE Transactions on Pattern Analysis and Machine Intelligence}, 44\penalty0 (12):\penalty0 9217--9220, 2022.
\newblock Conference Name: IEEE Transactions on Pattern Analysis and Machine Intelligence.

\bibitem[Allen et~al.(2022)Allen, St-Yves, Wu, Breedlove, Prince, Dowdle, Nau, Caron, Pestilli, Charest, et~al.]{allen2022massive}
Emily~J Allen, Ghislain St-Yves, Yihan Wu, Jesse~L Breedlove, Jacob~S Prince, Logan~T Dowdle, Matthias Nau, Brad Caron, Franco Pestilli, Ian Charest, et~al.
\newblock {A massive 7T fMRI dataset to bridge cognitive neuroscience and artificial intelligence}.
\newblock \emph{Nature neuroscience}, 25\penalty0 (1):\penalty0 116--126, 2022.

\bibitem[Bai et~al.(2023)Bai, Wang, Cao, Ge, Yuan, and Shan]{bai2023dreamdiffusion}
Yunpeng Bai, Xintao Wang, Yan-pei Cao, Yixiao Ge, Chun Yuan, and Ying Shan.
\newblock {Dreamdiffusion: Generating high-quality images from brain eeg signals}.
\newblock \emph{arXiv preprint arXiv:2306.16934}, 2023.

\bibitem[Benchetrit et~al.(2024)Benchetrit, Banville, and King]{benchetrit2024brain}
Yohann Benchetrit, Hubert Banville, and Jean-Remi King.
\newblock {{Brain decoding: toward real-time reconstruction of visual perception}}.
\newblock In \emph{{The Twelfth International Conference on Learning Representations}}, 2024.

\bibitem[Bomatter et~al.(2024)Bomatter, Paillard, Garces, Hipp, and Engemann]{Bomatter2023}
Philipp Bomatter, Joseph Paillard, Pilar Garces, J{\"o}rg Hipp, and Denis Engemann.
\newblock {{Machine learning of brain-specific biomarkers from EEG}}.
\newblock \emph{bioRxiv}, 2024.

\bibitem[Chang et~al.(2019)Chang, Pyles, Marcus, Gupta, Tarr, and Aminoff]{chang2019bold5000}
Nadine Chang, John~A Pyles, Austin Marcus, Abhinav Gupta, Michael~J Tarr, and Elissa~M Aminoff.
\newblock {BOLD5000, a public fMRI dataset while viewing 5000 visual images}.
\newblock \emph{Scientific data}, 6\penalty0 (1):\penalty0 49, 2019.

\bibitem[Chen et~al.(2023)Chen, Xu, Qing, Li, and Zhou]{chenstructure}
Zijiao Chen, Jonathan Xu, Jiaxin Qing, Ruilin Li, and Juan~Helen Zhou.
\newblock {Structure-Preserved Image Reconstruction from Brain Recordings}.
\newblock In preparation, 2023.

\bibitem[Chen et~al.(2024)Chen, Qing, and Zhou]{chen2024cinematic}
Zijiao Chen, Jiaxin Qing, and Juan~Helen Zhou.
\newblock {Cinematic mindscapes: High-quality video reconstruction from brain activity}.
\newblock \emph{Advances in Neural Information Processing Systems}, 36, 2024.

\bibitem[Deng et~al.(2009)Deng, Dong, Socher, Li, Li, and Fei-Fei]{deng2009imagenet}
Jia Deng, Wei Dong, Richard Socher, Li-Jia Li, Kai Li, and Li Fei-Fei.
\newblock Imagenet: A large-scale hierarchical image database.
\newblock In \emph{2009 IEEE conference on computer vision and pattern recognition}, pages 248--255. Ieee, 2009.

\bibitem[Dijkstra et~al.(2018)Dijkstra, Mostert, Lange, Bosch, and van Gerven]{dijkstra2018differential}
Nadine Dijkstra, Pim Mostert, Floris P~de Lange, Sander Bosch, and Marcel~AJ van Gerven.
\newblock {Differential temporal dynamics during visual imagery and perception}.
\newblock \emph{Elife}, 7:\penalty0 e33904, 2018.

\bibitem[Gifford et~al.(2022)Gifford, Dwivedi, Roig, and Cichy]{gifford_large_2022}
Alessandro~T. Gifford, Kshitij Dwivedi, Gemma Roig, and Radoslaw~M. Cichy.
\newblock {A large and rich {EEG} dataset for modeling human visual object recognition}.
\newblock \emph{NeuroImage}, 264:\penalty0 119754, 2022.

\bibitem[Gramfort et~al.(2013)Gramfort, Luessi, Larson, Engemann, Strohmeier, Brodbeck, Goj, Jas, Brooks, Parkkonen, et~al.]{gramfort2013meg}
Alexandre Gramfort, Martin Luessi, Eric Larson, Denis~A Engemann, Daniel Strohmeier, Christian Brodbeck, Roman Goj, Mainak Jas, Teon Brooks, Lauri Parkkonen, et~al.
\newblock {MEG and EEG data analysis with MNE-Python}.
\newblock \emph{Frontiers in neuroscience}, 7:\penalty0 70133, 2013.

\bibitem[Grootswagers et~al.(2019)Grootswagers, Robinson, and Carlson]{grootswagers2019representational}
Tijl Grootswagers, Amanda~K Robinson, and Thomas~A Carlson.
\newblock {The representational dynamics of visual objects in rapid serial visual processing streams}.
\newblock \emph{NeuroImage}, 188:\penalty0 668--679, 2019.

\bibitem[Grootswagers et~al.(2022)Grootswagers, Zhou, Robinson, Hebart, and Carlson]{grootswagers2022human}
Tijl Grootswagers, Ivy Zhou, Amanda~K Robinson, Martin~N Hebart, and Thomas~A Carlson.
\newblock {Human EEG recordings for 1,854 concepts presented in rapid serial visual presentation streams}.
\newblock \emph{Scientific Data}, 9\penalty0 (1):\penalty0 3, 2022.

\bibitem[Harel et~al.(2016)Harel, Groen, Kravitz, Deouell, and Baker]{harel2016temporal}
Assaf Harel, Iris~IA Groen, Dwight~J Kravitz, Leon~Y Deouell, and Chris~I Baker.
\newblock {The temporal dynamics of scene processing: A multifaceted EEG investigation}.
\newblock \emph{Eneuro}, 3\penalty0 (5), 2016.

\bibitem[Hyv{\"a}rinen et~al.(2001)Hyv{\"a}rinen, Karhunen, and Oja]{hyvarinen2001independent}
Aapo Hyv{\"a}rinen, Juha Karhunen, and Erkki Oja.
\newblock {Independent component analysis, adaptive and learning systems for signal processing, communications, and control}.
\newblock \emph{John Wiley \& Sons, Inc}, 1:\penalty0 11--14, 2001.

\bibitem[Jas et~al.(2017)Jas, Engemann, Bekhti, Raimondo, and Gramfort]{autoreject2017jas}
Mainak Jas, Denis~A Engemann, Yousra Bekhti, Federico Raimondo, and Alexandre Gramfort.
\newblock {Autoreject: Automated artifact rejection for MEG and EEG data}.
\newblock \emph{NeuroImage}, 159:\penalty0 417--429, 2017.

\bibitem[Jayaram and Barachant(2018)]{jayaram2018moabb}
Vinay Jayaram and Alexandre Barachant.
\newblock {{MOABB: trustworthy algorithm benchmarking for BCIs}}.
\newblock \emph{Journal of neural engineering}, 15\penalty0 (6):\penalty0 066011, 2018.

\bibitem[Kamitani(2017)]{horikawa2017god}
Tomoyasu Horikawa \&~Yukiyasu Kamitani.
\newblock {Generic decoding of seen and imagined objects using hierarchical visual features}.
\newblock \emph{Nature Communications}, 2017.

\bibitem[Kavasidis et~al.(2017)Kavasidis, Palazzo, Spampinato, Giordano, and Shah]{kavasidis_brain2image_2017}
Isaak Kavasidis, Simone Palazzo, Concetto Spampinato, Daniela Giordano, and Mubarak Shah.
\newblock {\textit{{Brain2Image}}: {Converting} {Brain} {Signals} into {Images}}.
\newblock In \emph{{Proceedings of the 25th {ACM} international conference on {Multimedia}}}, pages 1809--1817, Mountain View California USA, 2017. ACM.

\bibitem[Khaleghi et~al.(2022)Khaleghi, Rezaii, Beheshti, Meshgini, Sheykhivand, and Danishvar]{khaleghi_visual_2022}
Nastaran Khaleghi, Tohid~Yousefi Rezaii, Soosan Beheshti, Saeed Meshgini, Sobhan Sheykhivand, and Sebelan Danishvar.
\newblock {Visual {Saliency} and {Image} {Reconstruction} from {EEG} {Signals} via an {Effective} {Geometric} {Deep} {Network}-{Based} {Generative} {Adversarial} {Network}}.
\newblock \emph{Electronics}, 11\penalty0 (21):\penalty0 3637, 2022.
\newblock Number: 21 Publisher: Multidisciplinary Digital Publishing Institute.

\bibitem[King et~al.(2020)King, Gwilliams, Holdgraf, Sassenhagen, Barachant, Engemann, Larson, and Gramfort]{king:2020}
Jean-Rémi King, Laura Gwilliams, Chris Holdgraf, Jona Sassenhagen, Alexandre Barachant, Denis Engemann, Eric Larson, and Alexandre Gramfort.
\newblock {{{Encoding and Decoding Framework to Uncover the Algorithms of Cognition}}}.
\newblock In \emph{{The Cognitive Neurosciences}}. The MIT Press, 2020.

\bibitem[Ku(2018)]{ku2018selective}
Yixuan Ku.
\newblock Selective attention on representations in working memory: cognitive and neural mechanisms.
\newblock \emph{PeerJ}, 6:\penalty0 e4585, 2018.

\bibitem[Lan et~al.(2023)Lan, Ren, Wang, Zheng, Li, Lu, and Qiu]{lan_seeing_2023}
Yu-Ting Lan, Kan Ren, Yansen Wang, Wei-Long Zheng, Dongsheng Li, Bao-Liang Lu, and Lili Qiu.
\newblock {Seeing through the {Brain}: {Image} {Reconstruction} of {Visual} {Perception} from {Human} {Brain} {Signals}}, 2023.
\newblock arXiv:2308.02510 [cs, eess, q-bio].

\bibitem[Le et~al.(2021)Le, Ambrogioni, Seeliger, G{\"u}{\c{c}}l{\"u}t{\"u}rk, Van~Gerven, and G{\"u}{\c{c}}l{\"u}]{le2021brain2pix}
Lynn Le, Luca Ambrogioni, Katja Seeliger, Ya{\u{g}}mur G{\"u}{\c{c}}l{\"u}t{\"u}rk, Marcel Van~Gerven, and Umut G{\"u}{\c{c}}l{\"u}.
\newblock Brain2pix: Fully convolutional naturalistic video reconstruction from brain activity.
\newblock \emph{BioRxiv}, pages 2021--02, 2021.

\bibitem[Li et~al.(2021)Li, Johansen, Ahmed, Ilyevsky, Wilbur, Bharadwaj, and Siskind]{9264220}
Ren Li, Jared~S. Johansen, Hamad Ahmed, Thomas~V. Ilyevsky, Ronnie~B. Wilbur, Hari~M. Bharadwaj, and Jeffrey~Mark Siskind.
\newblock The perils and pitfalls of block design for eeg classification experiments.
\newblock \emph{IEEE Transactions on Pattern Analysis and Machine Intelligence}, 43\penalty0 (1):\penalty0 316--333, 2021.

\bibitem[Lin et~al.(2014{\natexlab{a}})Lin, Maire, Belongie, Hays, Perona, Ramanan, Doll{\'a}r, and Zitnick]{coco2014}
Tsung-Yi Lin, Michael Maire, Serge Belongie, James Hays, Pietro Perona, Deva Ramanan, Piotr Doll{\'a}r, and C~Lawrence Zitnick.
\newblock {Microsoft coco: Common objects in context}.
\newblock In \emph{{Computer Vision--ECCV 2014: 13th European Conference, Zurich, Switzerland, September 6-12, 2014, Proceedings, Part V 13}}, pages 740--755. Springer, 2014{\natexlab{a}}.

\bibitem[Lin et~al.(2014{\natexlab{b}})Lin, Maire, Belongie, Hays, Perona, Ramanan, Doll{\'a}r, and Zitnick]{lin2014microsoft}
Tsung-Yi Lin, Michael Maire, Serge Belongie, James Hays, Pietro Perona, Deva Ramanan, Piotr Doll{\'a}r, and C~Lawrence Zitnick.
\newblock {Microsoft coco: Common objects in context}.
\newblock In \emph{{Computer Vision--ECCV 2014: 13th European Conference, Zurich, Switzerland, September 6-12, 2014, Proceedings, Part V 13}}, pages 740--755. Springer, 2014{\natexlab{b}}.

\bibitem[Luck and Kappenman(2020)]{luck2020blog}
Steven~J Luck and Emily Kappenman.
\newblock A new metric for quantifying erp data quality.
\newblock \url{https://erpinfo.org/blog/2020/4/28/data-quality}, 2020.

\bibitem[Luck et~al.(2021)Luck, Stewart, Simmons, and Rhemtulla]{luck2021standardized}
Steven~J Luck, Andrew~X Stewart, Aaron~Matthew Simmons, and Mijke Rhemtulla.
\newblock Standardized measurement error: A universal metric of data quality for averaged event-related potentials.
\newblock \emph{Psychophysiology}, 58\penalty0 (6):\penalty0 e13793, 2021.

\bibitem[Mishra et~al.(2023)Mishra, Sharma, Jha, and Bhavsar]{mishra_neurogan_2023}
Rahul Mishra, Krishan Sharma, R.~R. Jha, and Arnav Bhavsar.
\newblock {{NeuroGAN}: image reconstruction from {EEG} signals via an attention-based {GAN}}.
\newblock \emph{Neural Computing and Applications}, 35\penalty0 (12):\penalty0 9181--9192, 2023.

\bibitem[Nemrodov et~al.(2018)Nemrodov, Niemeier, Patel, and Nestor]{nemrodov2018neural}
Dan Nemrodov, Matthias Niemeier, Ashutosh Patel, and Adrian Nestor.
\newblock {The neural dynamics of facial identity processing: insights from EEG-based pattern analysis and image reconstruction}.
\newblock \emph{Eneuro}, 5\penalty0 (1), 2018.

\bibitem[Nemrodov et~al.(2020)Nemrodov, Ling, Nudnou, Roberts, Cant, Lee, and Nestor]{nemrodov2020multivariate}
Dan Nemrodov, Shouyu Ling, Ilya Nudnou, Tyler Roberts, Jonathan~S. Cant, Andy C.~H. Lee, and Adrian Nestor.
\newblock {A multivariate investigation of visual word, face, and ensemble processing: Perspectives from EEG-based decoding and feature selection}.
\newblock \emph{Psychophysiology}, 57\penalty0 (3):\penalty0 e13511, 2020.

\bibitem[Radford et~al.(2021)Radford, Kim, Hallacy, Ramesh, Goh, Agarwal, Sastry, Askell, Mishkin, Clark, et~al.]{radford2021learning}
Alec Radford, Jong~Wook Kim, Chris Hallacy, Aditya Ramesh, Gabriel Goh, Sandhini Agarwal, Girish Sastry, Amanda Askell, Pamela Mishkin, Jack Clark, et~al.
\newblock Learning transferable visual models from natural language supervision.
\newblock In \emph{International conference on machine learning}, pages 8748--8763. PMLR, 2021.

\bibitem[Roy et~al.(2019)Roy, Banville, Albuquerque, Gramfort, Falk, and Faubert]{Roy2019}
Yannick Roy, Hubert Banville, Isabela Albuquerque, Alexandre Gramfort, Tiago~H Falk, and Jocelyn Faubert.
\newblock {{Deep learning-based electroencephalography analysis: a systematic review}}.
\newblock \emph{Journal of Neural Engineering}, 16\penalty0 (5):\penalty0 051001, 2019.

\bibitem[Scotti et~al.(2023)Scotti, Banerjee, Goode, Shabalin, Nguyen, Ethan, Dempster, Verlinde, Yundler, Weisberg, Norman, and Abraham]{scotti2023reconstructing}
Paul~Steven Scotti, Atmadeep Banerjee, Jimmie Goode, Stepan Shabalin, Alex Nguyen, Cohen Ethan, Aidan~James Dempster, Nathalie Verlinde, Elad Yundler, David Weisberg, Kenneth Norman, and Tanishq~Mathew Abraham.
\newblock {{Reconstructing the Mind's Eye: f{MRI}-to-Image with Contrastive Learning and Diffusion Priors}}.
\newblock In \emph{{Thirty-seventh Conference on Neural Information Processing Systems}}, 2023.

\bibitem[Scotti et~al.(2024)Scotti, Tripathy, Villanueva, Kneeland, Chen, Narang, Santhirasegaran, Xu, Naselaris, Norman, et~al.]{scotti2024mindeye}
Paul~S Scotti, Mihir Tripathy, Cesar Kadir~Torrico Villanueva, Reese Kneeland, Tong Chen, Ashutosh Narang, Charan Santhirasegaran, Jonathan Xu, Thomas Naselaris, Kenneth~A Norman, et~al.
\newblock {{MindEye2: Shared-Subject Models Enable fMRI-To-Image With 1 Hour of Data}}.
\newblock \emph{arXiv preprint arXiv:2403.11207}, 2024.

\bibitem[Singh et~al.(2023)Singh, Pandey, Miyapuram, and Raman]{singh_eeg2image_2023}
Prajwal Singh, Pankaj Pandey, Krishna Miyapuram, and Shanmuganathan Raman.
\newblock {{EEG2IMAGE}: {Image} {Reconstruction} from {EEG} {Brain} {Signals}}, 2023.
\newblock arXiv:2302.10121 [cs, q-bio].

\bibitem[Singh et~al.(2024)Singh, Dalal, Vashishtha, Miyapuram, and Raman]{singh2024learning}
Prajwal Singh, Dwip Dalal, Gautam Vashishtha, Krishna Miyapuram, and Shanmuganathan Raman.
\newblock {Learning Robust Deep Visual Representations from EEG Brain Recordings}.
\newblock In \emph{{Proceedings of the IEEE/CVF Winter Conference on Applications of Computer Vision}}, pages 7553--7562, 2024.

\bibitem[Spampinato et~al.(2017)Spampinato, Palazzo, Kavasidis, Giordano, Souly, and Shah]{spampinato2017deeplearning}
Concetto Spampinato, Simone Palazzo, Isaak Kavasidis, Daniela Giordano, Nasim Souly, and Mubarak Shah.
\newblock {Deep learning human mind for automated visual classification}.
\newblock In \emph{{Proceedings of the IEEE conference on computer vision and pattern recognition}}, pages 6809--6817, 2017.

\bibitem[Tanner et~al.(2016)Tanner, Norton, Morgan-Short, and Luck]{tanner2016high}
Darren Tanner, James~JS Norton, Kara Morgan-Short, and Steven~J Luck.
\newblock {On high-pass filter artifacts (they’re real) and baseline correction (it’sa good idea) in ERP/ERMF analysis}.
\newblock \emph{Journal of neuroscience methods}, 266:\penalty0 166--170, 2016.

\bibitem[Tao et~al.(2021)Tao, Sun, Muhamed, Genc, Jackson, Arsanjani, Yaddanapudi, Li, and Kumar]{tao2021gated}
Yunzhe Tao, Tao Sun, Aashiq Muhamed, Sahika Genc, Dylan Jackson, Ali Arsanjani, Suri Yaddanapudi, Liang Li, and Prachi Kumar.
\newblock {Gated transformer for decoding human brain EEG signals}.
\newblock In \emph{{2021 43rd Annual International Conference of the IEEE Engineering in Medicine \& Biology Society (EMBC)}}, pages 125--130. IEEE, 2021.

\bibitem[Teichmann et~al.(2023)Teichmann, Hebart, and Baker]{teichmann2023multidimensional}
Lina Teichmann, Martin~N Hebart, and Chris~I Baker.
\newblock {Multidimensional object properties are dynamically represented in the human brain}.
\newblock \emph{bioRxiv}, 2023.

\bibitem[Thorpe et~al.(1996)Thorpe, Fize, and Marlot]{thorpe1996speed}
Simon Thorpe, Denis Fize, and Catherine Marlot.
\newblock {Speed of processing in the human visual system}.
\newblock \emph{nature}, 381\penalty0 (6582):\penalty0 520--522, 1996.

\bibitem[Tirupattur et~al.(2018)Tirupattur, Rawat, Spampinato, and Shah]{tirupattur_thoughtviz_2018}
Praveen Tirupattur, Yogesh~Singh Rawat, Concetto Spampinato, and Mubarak Shah.
\newblock {{ThoughtViz}: {Visualizing} {Human} {Thoughts} {Using} {Generative} {Adversarial} {Network}}.
\newblock In \emph{{Proceedings of the 26th {ACM} international conference on {Multimedia}}}, pages 950--958, Seoul Republic of Korea, 2018. ACM.

\bibitem[Vivancos and Cuesta(2022)]{vivancos2022mindbigdata}
David Vivancos and Felix Cuesta.
\newblock {MindBigData 2022 A Large Dataset of Brain Signals}.
\newblock \emph{arXiv preprint arXiv:2212.14746}, 2022.

\bibitem[Wakita et~al.(2021)Wakita, Orima, and Motoyoshi]{wakita_photorealistic_2021}
Suguru Wakita, Taiki Orima, and Isamu Motoyoshi.
\newblock {Photorealistic {Reconstruction} of {Visual} {Texture} {From} {EEG} {Signals}}.
\newblock \emph{Frontiers in Computational Neuroscience}, 15, 2021.

\bibitem[Yamins and DiCarlo(2016)]{yamins2016using}
Daniel~LK Yamins and James~J DiCarlo.
\newblock Using goal-driven deep learning models to understand sensory cortex.
\newblock \emph{Nature neuroscience}, 19\penalty0 (3):\penalty0 356--365, 2016.

\end{thebibliography}
}


\end{document}